\documentstyle[twoside,psfig]{article}

\catcode`\@=11
\long\def\@makefntext#1{
\protect\noindent \hbox to 3.2pt {\hskip-.9pt  
$^{{\eightrm\@thefnmark}}$\hfil}#1\hfill}		

\def\thefootnote{\fnsymbol{footnote}}
\def\@makefnmark{\hbox to 0pt{$^{\@thefnmark}$\hss}}	
	
\def\ps@myheadings{\let\@mkboth\@gobbletwo
\def\@oddhead{\hbox{}
\rightmark\hfil\eightrm\thepage}   
\def\@oddfoot{}\def\@evenhead{\eightrm\thepage\hfil
\leftmark\hbox{}}\def\@evenfoot{}
\def\sectionmark##1{}\def\subsectionmark##1{}}

\oddsidemargin=\evensidemargin
\renewcommand{\thefootnote}{\fnsymbol{footnote}}

\newcounter{sectionc}\newcounter{subsectionc}\newcounter{subsubsectionc}
\renewcommand{\section}[1] {\vspace{12pt}\addtocounter{sectionc}{1} 
\setcounter{subsectionc}{0}\setcounter{subsubsectionc}{0}\noindent 
	{\tenbf\thesectionc. #1}\par\vspace{5pt}}
\renewcommand{\subsection}[1] {\vspace{12pt}\addtocounter{subsectionc}{1} 
	\setcounter{subsubsectionc}{0}\noindent 
	{\bf\thesectionc.\thesubsectionc. {\kern1pt \bfit #1}}\par\vspace{5pt}}
\renewcommand{\subsubsection}[1] {\vspace{12pt}\addtocounter{subsubsectionc}{1}
	\noindent{\tenrm\thesectionc.\thesubsectionc.\thesubsubsectionc.
	{\kern1pt \tenit #1}}\par\vspace{5pt}}
\newcommand{\nonumsection}[1] {\vspace{12pt}\noindent{\tenbf #1}
	\par\vspace{5pt}}

\newcounter{appendixc}
\newcounter{subappendixc}[appendixc]
\newcounter{subsubappendixc}[subappendixc]
\renewcommand{\thesubappendixc}{\Alph{appendixc}.\arabic{subappendixc}}
\renewcommand{\thesubsubappendixc}
	{\Alph{appendixc}.\arabic{subappendixc}.\arabic{subsubappendixc}}

\renewcommand{\appendix}[1] {\vspace{12pt}
        \refstepcounter{appendixc}
        \setcounter{figure}{0}
        \setcounter{table}{0}
        \setcounter{lemma}{0}
        \setcounter{theorem}{0}
        \setcounter{corollary}{0}
        \setcounter{definition}{0}
        \setcounter{equation}{0}
        \renewcommand{\thefigure}{\Alph{appendixc}.\arabic{figure}}
        \renewcommand{\thetable}{\Alph{appendixc}.\arabic{table}}
        \renewcommand{\theappendixc}{\Alph{appendixc}}
        \renewcommand{\thelemma}{\Alph{appendixc}.\arabic{lemma}}
        \renewcommand{\thetheorem}{\Alph{appendixc}.\arabic{theorem}}
        \renewcommand{\thedefinition}{\Alph{appendixc}.\arabic{definition}}
        \renewcommand{\thecorollary}{\Alph{appendixc}.\arabic{corollary}}
        \renewcommand{\theequation}{\Alph{appendixc}.\arabic{equation}}
        \noindent{\tenbf Appendix \theappendixc #1}\par\vspace{5pt}}
\newcommand{\subappendix}[1] {\vspace{12pt}
        \refstepcounter{subappendixc}
        \noindent{\bf Appendix \thesubappendixc. {\kern1pt \bfit #1}}
	\par\vspace{5pt}}
\newcommand{\subsubappendix}[1] {\vspace{12pt}
        \refstepcounter{subsubappendixc}
        \noindent{\rm Appendix \thesubsubappendixc. {\kern1pt \tenit #1}}
	\par\vspace{5pt}}

\newcommand{\textlineskip}{\baselineskip=13pt}
\newcommand{\smalllineskip}{\baselineskip=10pt}

\def\eightcirc{
\begin{picture}(0,0)
\put(4.4,1.8){\circle{6.5}}
\end{picture}}
\def\eightcopyright{\eightcirc\kern2.7pt\hbox{\eightrm c}} 

\newcommand{\copyrightheading}[1]
	{\vspace*{-2.5cm}\smalllineskip{\flushleft
	{\footnotesize Modern Physics Letters A, #1}\\
	{\footnotesize $\eightcopyright$\, World Scientific Publishing
	 Company}\\
	 }}

\newcommand{\publisher}[2]{{\begin{center}\footnotesize\smalllineskip 
	Received #1\\
	Revised #2
	\end{center}
	}}

\def\abstracts#1#2#3{{
	\centering{\begin{minipage}{4.5in}\footnotesize\baselineskip=10pt
	\parindent=0pt #1\par 
	\parindent=15pt #2\par
	\parindent=15pt #3
	\end{minipage}}\par}} 

\renewenvironment{thebibliography}[1]
	{\frenchspacing
	 \ninerm\baselineskip=11pt
	 \begin{list}{\arabic{enumi}.}
        {\usecounter{enumi}\setlength{\parsep}{0pt}     
	 \setlength{\leftmargin 12.7pt}{\rightmargin 0pt} 
         \setlength{\itemsep}{0pt} \settowidth
	{\labelwidth}{#1.}\sloppy}}{\end{list}}

\newcounter{itemlistc}
\newcounter{romanlistc}
\newcounter{alphlistc}
\newcounter{arabiclistc}

\newcommand{\fcaption}[1]{
        \refstepcounter{figure}
        \setbox\@tempboxa = \hbox{\footnotesize Fig.~\thefigure. #1}
        \ifdim \wd\@tempboxa > 5in
           {\begin{center}
        \parbox{5in}{\footnotesize\smalllineskip Fig.~\thefigure. #1}
            \end{center}}
        \else
             {\begin{center}
             {\footnotesize Fig.~\thefigure. #1}
              \end{center}}
        \fi}

\newcommand{\tcaption}[1]{
        \refstepcounter{table}
        \setbox\@tempboxa = \hbox{\footnotesize Table~\thetable. #1}
        \ifdim \wd\@tempboxa > 5in
           {\begin{center}
        \parbox{5in}{\footnotesize\smalllineskip Table~\thetable. #1}
            \end{center}}
        \else
             {\begin{center}
             {\footnotesize Table~\thetable. #1}
              \end{center}}
        \fi}

\def\@citex[#1]#2{\if@filesw\immediate\write\@auxout
	{\string\citation{#2}}\fi
\def\@citea{}\@cite{\@for\@citeb:=#2\do
	{\@citea\def\@citea{,}\@ifundefined
	{b@\@citeb}{{\bf ?}\@warning
	{Citation `\@citeb' on page \thepage \space undefined}}
	{\csname b@\@citeb\endcsname}}}{#1}}

\newif\if@cghi
\def\cite{\@cghitrue\@ifnextchar [{\@tempswatrue
	\@citex}{\@tempswafalse\@citex[]}}
\def\citelow{\@cghifalse\@ifnextchar [{\@tempswatrue
	\@citex}{\@tempswafalse\@citex[]}}
\def\@cite#1#2{{$\null^{#1}$\if@tempswa\typeout
	{IJCGA warning: optional citation argument 
	ignored: `#2'} \fi}}

\def\pmb#1{\setbox0=\hbox{#1}
	\kern-.025em\copy0\kern-\wd0
	\kern.05em\copy0\kern-\wd0
	\kern-.025em\raise.0433em\box0}

\def\fnt#1#2{\footnotetext{\kern-.3em
	{$^{\mbox{\scriptsize #1}}$}{#2}}}

\def\fpage#1{\begingroup
\voffset=.3in
\thispagestyle{empty}\begin{table}[b]\centerline{\footnotesize #1}
	\end{table}\endgroup}

\font\tenrm=cmr10
\font\tenit=cmti10 
\font\tenbf=cmbx10
\font\bfit=cmbxti10 at 10pt
\font\ninerm=cmr9

\font\eightrm=cmr8

\def\qed{\hbox{${\vcenter{\vbox{			
   \hrule height 0.4pt\hbox{\vrule width 0.4pt height 6pt
   \kern5pt\vrule width 0.4pt}\hrule height 0.4pt}}}$}}

\renewcommand{\thefootnote}{\fnsymbol{footnote}}	

\begin{document}
\setlength{\textheight}{7.7truein}  


\normalsize\textlineskip
\thispagestyle{empty}
\setcounter{page}{1}

\copyrightheading{}			

\vspace*{0.88truein}

\fpage{1}
\centerline{{\bf Geometry of theory space and RG flows}
\footnote{\em Invited talk at the First IUCAA Workshop on `Interface of
Gravitational and Quantum Realms, IUCAA, Pune, Dec 17-21, 2001}}
\baselineskip=13pt
\vspace*{0.37truein}
\baselineskip=12pt
\centerline{\footnotesize Sayan Kar }

\centerline{\footnotesize  Department of Physics and Centre for
Theoretical Studies,}
\centerline{\footnotesize Indian Institute of Technology}
\centerline{\footnotesize  Kharagpur
721 302, India} 

\baselineskip=10pt
\vspace*{10pt}

\publisher{(received date)}{(revised date)}

\vspace*{0.21truein}
\abstracts{The space of couplings of a given theory is the arena of
interest in this article. Equipped with a metric ansatz akin to the
Fisher information matrix in the space of parameters in 
statistics (similar metrics in
physics are the Zamolodchikov metric or the O'Connor--Stephens metric)
we investigate the geometry of theory space through a study of 
specific examples. We then look into renormalisation group flows
in theory space and make an
attempt to characterise such flows via its isotropic expansion, rotation and
shear. Consequences arising from the evolution equation for the
isotropic expansion are discussed. 
We conclude by pointing out generalisations   
and pose some open questions.
}{}{}



\vspace*{1pt}\textlineskip	

\setcounter{footnote}{0}
\renewcommand{\thefootnote}{\alph{footnote}}

\section{Introduction}
\noindent
The use of a
language, different from the standard, may sometimes shed new light on
topics which do seem to be well-understood otherwise. A 
text-book example is the case of Maxwell's equations written in
the usual three-vector version and its reformulation using
Lorentz four vectors and the field tensor. In this article, we
intend to adopt a somewhat similar point of view. Usually 
a theory and its consequences are derived by an analysis 
in coordinate (or momentum) space. Here, we wish to
take a look via the  `theory space', 
which, essentially, is the space of the couplings which appear
in a theory with diverse interactions. We endow this space
with a line element (a metric tensor) through a prescription 
widely used in the theory of estimation in statistics {\cite{amari}}and also
discussed in the context of physics by various authors 
in different contexts in the past{\cite{zamolo} \cite{oconnor} \cite{frieden}}.   
Equipped with this metric ansatz we proceed towards understanding 
the geometry
of theory space, the nature of the flow of couplings (via finite
renormalisations) therein and make an attempt towards arriving at 
interesting, generic, theory--independent statements. Towards the end,
we pose a few open questions.  

Let us begin with simple examples {\cite{amari}} in the context
of statistics. This will help us in understanding the method
to be proposed for constructing a metric in the space of
couplings, later on.  We consider a Gaussian 
probability distribution $P(x, g)$ for a random variable $x$,
 parametrised by its mean $\mu$ 
and 
standard deviation, $\sigma$. $g$ is a compact notation for the
set of all parameters that appear in $P$ (here $g \equiv
 \{\mu,\sigma\}$)
We therefore have :

\begin{equation}
P(x,g) = \frac{1}{\sqrt{2\pi\sigma^2}} \exp{-\frac{(x-\mu)^2}{2\sigma^2}}
\end{equation}

We may now ask a set of questions : 

{\sf If $\left\{\mu,\sigma \right \}$ can be
thought as coordinates for a given space then :

(a) how do we define a 
metric in this space? 

(b) what are the features of this metric? 

(c)
can we make precise statements about the nature of the family of
probability distributions $P(x,g)$ with differing 
$\mu$ and $\sigma$ values by looking at the
metric and/or its properties?}

Let us now try to answer these questions.

Defining a new quantity $w = \ln P$, we now write down an expression
for the metric in the space of parameters. Known in the literature as
the Fisher information matrix, it is given as :

\begin{equation}
g_{ab}(g) = \int_{x_1}^{x_2} (\partial_a\partial_b w) P(x,\theta) dx
\end{equation}

The indices $a$ run over the parameters (here $\mu$ and $\sigma$ )
appearing in the probability distribution. In the case of the
Gaussian we have :

\begin{equation}
ds^2 = \frac{1}{\sigma^2} \left [  d\mu^{2} + 2d\sigma^2 \right ]
\end{equation}

This is the line element on the hyperbolic plane and it has a constant
Ricci scalar ($R=-\frac{1}{2}$). The metric coefficients are dependent
only on $\sigma$ and a coordinate singularity exists at $\sigma \rightarrow
0$. A coordinate transformation on $\sigma$ can get rid of this singularity 
in the
line element. Keeping $\mu$ or $\sigma$ fixed we can obtain expressions for
the `distance' between probability distributions with varying $\sigma$ or
$\mu$ respectively. The distance, in a certain sense, is a measure of
the relative entropy--it could measure the increase/decrease 
of disorder as we
change the values of the coordinates in theory space.

The answer to the question in (c) above, however, is not entirely clear.
Strictly speaking, the ideal issue here is whether we can obtain
theory--independent statements from a general analysis in theory space.
 We
shall provide an example of such a statement in the penultimate
section of this article.
 
Examples, in the physical context, where a Gaussian distribution in the
form shown in Eq. (1) may arise
are abundant. Consider a linear harmonic oscillator in a constant electric
field. The probability distribution for the
ground state wave function can be easily recast into the form
(1). 
A detailed analysis of such quantum mechanical examples in different
contexts will be
discussed elsewhere {\cite{sk}}.

\section{Theory spaces in field theory}
\noindent
The probability distribution mentioned in (1), with the replacement
of coordinates by a field $\phi$ could be thought of as the
partition function of a massive, Euclidean,  scalar field theory in zero dimensions 
in the presence of an external interaction via a $j\phi$ coupling
($j$ being a constant). Extending to higher dimensions one can
set up the following framework {\cite{dolan97,dolan98}}.

We begin with the partition function in an Euclidean space of
$D$ dimensions ($g^a$ ($a=1,2 .... ,n$ denotes a set of n couplings) :

\begin{equation}
Z[g] = \int D\phi e^{-S[\phi]}
\end{equation}

It is convenient to define the quantity $W = -\ln Z$, such that :
\begin{equation}
\int D\phi e^{-S[\phi] + W} = 1
\end{equation}

This implies $dW =  \langle dS \rangle$ with $dW = \partial_aW dg^a$ and
$dS =\partial_a Sdg^a$. The O'Connor--Stephen's form of the metric in
the space of couplings ($g^a$ are n couplings which appear in the action
through the expression $S = S_0 + g^{a}\int {\bf \Phi_{a}}d^D x$, where ${\bf \Phi_a}$
in general could be composite operators) :

\begin{equation}
g_{ab} = \langle (dS-dW) \otimes (dS-dW)\rangle = -\partial_a\partial_b w
+\frac{1}{V}\langle \partial_a\partial_b S \rangle 
\end{equation} 

where $W=\int w d^Dx$ and $V$ is the spatial volume.

For an action with interaction terms linear in the couplings, we have 
$g_{ab} = -\partial_a\partial_b w$
Usually, this expression is employed in order to find the metric in the 
space of theories.

For theories where the partition function is integrable 
one can hope to obtain exact results. In cases where our only
analytic tool is perturbation theory we can only get the metric
in the space of couplings upto a corresponding order in 
the perturbation parameter. The latter case requires a 
renormalisable quantum field theory.

We now discuss the geometry of the theory space of a couple of
well-known theories. The examples involve partition functions which are
exactly integrable.
These serve as
useful illustrations of the formalism outlined above.

\subsection{Examples}

We first consider an Euclidean scalar field theory 
with a mass term and a $j\phi$ coupling 
in $D$ dimensions. The mass parameter $m$ is also defined as a `coupling'
in a generalised sense. The partition function is :

\begin{equation}
Z[j,m^2] = \int D\phi \exp\left[ -S(\phi;j,m^2)\right ]
\end{equation}

where $S(\phi;j,m^2) = \int d^Dx \left \{ \frac{1}{2} \phi \left (
-\nabla^2 + m^2 \right ) \phi +j\phi \right \} $.

Following the prescription outlined in the previous section it is
possible to write down an expression for the metric. 
This is given (for $D=1,2,3$) as :

\begin{equation}
ds^2 = dr^2 + r^{\frac{4}{D}}d\chi^2
\end{equation}

where $r$ and $\chi$ are defined as :

\begin{eqnarray}
\chi = 2\sqrt{\pi}\left [ \frac{D}{4}\sqrt{\frac{2}{\Gamma(2-\frac{D}{2})}
}\right ] ^{\frac{2}{D}} \xi \\ 
r = \frac{4}{D} \sqrt{\frac{\Gamma(2-\frac{D}{2})}{2}}\left (\frac{m^{2}}{4\pi}
\right ) ^{\frac{D}{4}}
\end{eqnarray}

The  Ricci scalar turns out to be ;

\begin{equation}
R= -\frac{2(2-D)}{D^{2}r^{2}}
\end{equation}

Separate calculations reveal, additionally, $R=-\frac{1}{2}$ 
(the hyperbolic plane) (for $D=0$ and
$R=0$ (for $D=4$). It is worth noting that $D=2$ is like a crossover point
in the sense that for $D<2$ $R<0$ while for $D=3$ $R>0$.

As we approach $r\rightarrow 0$ (for $D=1$ and $D=3$) the 
Ricci scalar diverges and we have a real singularity in the
metric. The limit $r\rightarrow 0$ also implies (via the
definition of $r$ provided above) $m^2\rightarrow 0$. A
singularity in the space of couplings seems to be related to
the massless feature of the theory, which, in turn, is 
linked to conformal invariance.  

It is also easy to note that for $m^{2}<0$ the metric in theory
space becomes Lorentzian. 

Our second example will be the Ising model in the presence of a
constant magnetic field.
The partition function is given as :

\begin{equation}
Z_{N} = \sum_{\{\sigma \}} \exp \left [K \sum_{j=1}^{N} \sigma_j \sigma_{j+1}
+ h \sum_{j=1}^{N} \sigma_j \right ]
\end{equation}

where $K=\frac{J}{kT}$ and $h=\frac{H}{kT}$. Using the pair of 
quantities $\rho = e^{2K}\sinh h$ and $K$ as the coordinates in the
space of couplings we can derive the line element. This is given as :

\begin{equation}
ds^2 = \frac{1}{\sqrt{1+\rho^2} e^{2K}\cosh h} \left [ \frac{4e^{4K}dK^2}
{\sqrt{1+\rho^2} + e^{2K}\cosh h} + \frac{d\rho^2}{1+\rho^2}\right ]
\end{equation}

Evaluating the Ricci scalar we find :

\begin{equation}
R = \frac{1}{2} \left [ 1 + \sqrt{\frac{e^{4K} + \rho^2}{1+\rho^2}}\right ]
\end{equation}

We note that as $K\rightarrow \infty$ (i.e. $T\rightarrow 0$, $R$ diverges).
The divergence of $R$ occurs as we approach the critical point $T\rightarrow 0$.

Other theories for which an analysis similar to the above has been carried
out include -- $\lambda \phi^4$ theory {\cite{dolan94}}, $O(N)$ model for
large $N$ {\cite{dolan98}}, $N=2$ supersymmetric Yang--Mills theory 
{\cite{dolan98plb}} and the Ising model on a Bethe lattice {\cite{dolan95}}.
It will certainly be worthwhile to work out more examples in order to
arrive at better insight into this approach.

\noindent

\section{Renormalisation group equation in theory space}
\noindent

In the above sections we have discussed the geometry of the space of couplings.
The method of endowing this space with a line element and ways to
calculate it have been outlined, with some simple examples. 
It is natural now to
investigate the nature of curves in this space--or, more importantly,
a family of curves. A curve in theory space represents a flow of
couplings. We are also familiar with such a notion of flow of couplings
in the theory of renormalisation in quantum field theory. 
Finite renormalisations generate
a flow of couplings and the set of such finite renormalisations are known to
form a semigroup, namely, the Renormalisation Group (RG). The flow of 
couplings, generated by the $\beta$-function vector field ($\beta^{a} =
\frac{dg^a}{d\kappa}$, $\kappa$ being the scale parameter which
parametrises points on the flow lines)  of a given 
field theory  
obeys the RG equation and is termed as a RG flow. We now embark on
a geometric analysis of RG flows in theory space.

Let us briefly recall how the RG equation arises.
We know that an
ambiguity arises in the choice of the infinite part of a regularised
Feynman amplitude. This necessitates the choice of a renormalisation scheme,
which,  
naturally, implies the existence of a scale or a renormalisation point.
The requirement
that physical amplitudes, say the one-particle irreducible amplitudes
, are independent of a choice of scale leads to the 
the RG equation.

Quantitatively, we can write the RG equation for the 2-point function as :

\begin{eqnarray}
\kappa\partial_{\kappa} \langle\Phi_{a}(x)\Phi_{b}(y)\rangle +
\beta^{c}\partial_{c} \langle\Phi_{a}(x)\Phi_{b}(y)\rangle +
\partial_{a}\beta^{c} \langle\Phi_{c}(x)\Phi_{b}(y)\rangle + \\ \nonumber
\partial_{b}\beta^{c} \langle\Phi_{a}(x)\Phi_{c}(y)\rangle 
=0
\end{eqnarray}

Using the facts that couplings are scaled to be dimensionless,
the $\Phi_{a}(x)$ have canonical mass dimension $D$ and the
scaling argument one can arrive at the equation :

\begin{eqnarray}
\left (x^{\mu}\frac{\partial}{\partial x^{\mu}} +
y^{\mu}\frac{\partial}{\partial y^{\mu}} \right )\langle\Phi_{a}(x)
\Phi_{b}(y)\rangle +2D \langle\Phi_{a}(x)\Phi_{b}(y)\rangle = \\ \nonumber
-\beta^{c}\partial_{c} \langle\Phi_{a}(x)\Phi_{b}(y)\rangle - 
\partial_{a}\beta^{c} \langle \Phi_{c}(x)\Phi_{b}(y)\rangle - 
\partial_{b}\beta^{c} \langle \Phi_{a}(x)\Phi_{c}(y)\rangle 
\end{eqnarray}

Integrating over all $y$ and using translational invariance we get

\begin{equation}
\beta^{c}\partial_{c}g_{ab} +(\partial_{a}\beta^{c})
g_{cb} + (\partial_{b}
\beta^{c} )g_{ca} = -Dg_{ab}
\end{equation}

which will finally yield (after some simple manipulations) :
  
\begin{equation}
\nabla_{a}\beta_{b} + \nabla_{b}\beta_{a} = - Dg_{ab}
\end{equation}

This is equivalent to the statement that RG flows are generated
by a $\beta$-function vector field which is conformally Killing.
This {\em geometric version} of the RG equation in theory space
is possible because of the equivalence :

\begin{equation}
{\cal L}_{\bf D} \Gamma (x_i, g^a fixed) \equiv 
{\cal L}_{\bf \beta}\Gamma (g^a,x_i
fixed)
\end{equation}

where $\cal L$ denotes a Lie derivative, $\bf D$ is the dilatation generator
and $\bf \beta$ is the above mentioned $\beta$--function vector 
field. This equivalence of descriptions was first noted by
Lassig {\cite{lassig}} and later elaborated in several papers by Dolan
{\cite{dolan94,dolan97}}.

\section{Expansion, rotation and shear of geodesic RG flows {\cite{sayan}}}

We now move on to a special class of RG flows which are 
geodesic. For theories in two dimensions such flows are possible.
Our results in this section are however extendible to non-geodesic
flows as well.

If a flow is geodesic as well as an RG flow it has to obey 
the RG equation as well as the geodesic equation $\beta^{a}\nabla_a\beta^b =0$
(for affinely parametrised geodesics). 
In such a case, the following decomposition of the covariant gradient
of a normalised ${\hat \beta}_{b} = \frac{\beta_b}{\sqrt{\beta_a\beta^a}}$ holds :

\begin{equation}
\nabla_a {\hat\beta}_b = \sigma_{ab} + \omega_{ab} + \frac{1}{n-1}h_{ab}\theta 
\end{equation}

where $h_{ab} = g_{ab} - {\hat \beta}_{a}{\hat \beta}_{b}$ is the 
projection tensor. $\sigma_{ab}$,
$\omega_{ab}$ and $\theta$ are the shear, rotation and isotropic expansion
for the geodesic flow under consideration. 
This way of analysing flows is traditionally employed in the context of
Riemannian geometry and General Relativity {\cite{wald}}, {\cite{hawk}}.
Shear, rotation and expansion are measures of the shape of the cross--sectional
area enclosing a geodesic congruence. Each of these quantities are functions
of the affine parameter $\lambda$ and are defined at every point on the flow.
A circle going over to a concentric but larger/smaller circle is
measured through the isotropic expansion $\theta$, Circles enclosing a
congruence, if they deform into ellipses would imply the existence of
a non--zero shear. A twist in the family of geodesics, much in the same
way as a twist on a rope made out of threads incorporates a rotation in the
flow.  

Using this decomposition in the
conformal Killing condition it is easy to note that the isotropic expansion
is uniquely determined in terms of the norm of the $\beta$-function 
vector field.

\begin{equation}
\theta = -\frac{D(n-1)}{2\beta}
\end{equation}

where $\beta = \sqrt{\beta_a \beta^a}$.

The negative sign here ensures a convergence of the flow towards a focal
point where all $\beta^{a}$ approach zero. In the language of field theory
such a point in coupling space is termed as a fixed point and implies a
conformally invariant theory defined through the values of the couplings
at that point. The crucial fact here is that this result is arrived at
without reference to any specific theory.

The shear and rotation of a geodesic RG flow can also be defined 
by taking the symmetric traceless and the antisymmetric parts of 
$\nabla_{a}{\hat\beta_b}$. 

Furthermore, following the approach of analysis employed in the context of
Riemannian geometry and General Relativity we may look into the evolution
equations for expansion, rotation and shear along the flow. In particular,
let us investigate the equation for the isotropic expansion, known in the
literature as the Raychaudhuri equation. This is given as:

\begin{equation}
\frac{d\theta}{d\lambda} + \frac{1}{n-1}\theta^2 +\sigma^2 -\omega^2
=-R_{ab}{{\hat\beta}^a}{{\hat\beta}^b}
\end{equation}

Setting $\sigma_{ab}$ and $\omega_{ab}$ equal to zero (which, incidentally,
is a consistent solution for the evolution equations for these quantities)
and using the expression for $\theta$ obtained using the conformal
Killing condition (RG equation) we obtain :

\begin{equation}
\frac{D(n-1)}{2}\left [ \frac{d\beta}{d\lambda} + \frac{D}{2} \right ]
= -R_{ab}\beta^{a}\beta^{b} 
\end{equation} 

Qualitative analysis leads to the following statement :  
The l.h.s. of this equation is a finite quantity. However, the
r.h.s. may diverge if $R_{ab}$ diverges (this may be possible if we
have a real singularity in theory space). In order to maintain the
finiteness of the l.h.s we therefore require the divergence in $R_{ab}$
be cancelled by a zero in the $\beta^{a}$. In specific types of
theory spaces such as an Einstein space ($R_{ab} =\Lambda g_{ab}$) or
a two dimensional one ($n=2$) it may also be possible to integrate
the above equation and obtain explicit solutions for the norm $\beta$
of the $\beta$--function vector field. 

It may also be noted that the conclusion about the focusing of any
geodesic RG flow towards a fixed point can also be generalised to
the case of non-geodesic flows. 
 
\noindent

\section{Outlook}

The
analysis discussed above holds for couplings which are not functions of the
coordinates. We may generalise our formalism to include such
options {\cite{erdmenger}}. Simplistic extensions of the models 
discussed above with the corresponding couplings now dependent on spacetime 
coordinates can be worked out as examples. However, this extension
becomes a necessity for the problem we propose below.

We would like to do this analysis for a nonlinear sigma model
coupled to gravity described by the usual Nambu--Goto or
Polyakov action for bosonic strings. Extensions to superstrings
is the logical follow-up but certainly not easy to carry out.  
Treating the metric coefficients as couplings (an attitude
primarily adopted in string theory) 
the {\em set} of all couplings becomes a space of functions
--an infinite dimensional space. Proceeding as before, we can 
define a metric in this space of all `real' metrics and use it to
study the `geometry' in this `superspace'(space of all metrics
{\em a la} Wheeler) . As before, we might
want to look at the $\beta$--functionals--find the nature of the
flows and check out the conclusions related to geodesic focusing of
the `trajectories' in theory space. Ofcourse, the focal point, in this case,
will occur for a certain `metric function'
which will imply a particular spacetime geometry. At the focal point
we therefore may have three notable features--(i) divergence of the
theory space Ricci tensor (ii) zero of the $\beta$-functional 
(iii) $\theta$ tending to $-\infty$. 

The usual analysis 
carried out to obtain low-energy effective actions for
string theories of various types proceeds by calculating
$\beta$--functionals and setting them to zero, in order to maintain
quantum conformal invariance {\cite{string}}. The equations of motion thus
obtained,
by setting the $\beta$--functionals to zero turn out to be
similar to the Einstein field equations of General Relativity, 
with some modifications.
Solving these equations one obtains line elements with a variety
of features (eg. black holes, naked singularities, cosmological models
and so on). In the language of strings, these line elements are the
admissible backgrounds in which a string can propagate. 

However, as far as we are aware, an explicit construction of the theory space 
, the geometry of RG flows, expressions for $\theta$, $\sigma_{ab}$ and 
$\omega_{ab}$ have not been attempted anywhere in the literature. 
To begin an analysis of this type one might  
specialise to the class of cosmological metrics (motivated by the
standard minisuperspace 
constructions)
characterised by $a(t)$ (the scale factor) and another field, say the
dilaton $\phi(t)$. The theory space here would be $\left \{ a(t),\phi(t) \right \}$
--essentially an infinite dimensional function space. In {\cite{tseyt:hepth}}
a discussion partially along the lines mentioned above was indeed carried out
though with somewhat different motivation. It will be worthwhile to
work out the theory space metric and the geometry of RG flows 
for this simplistic minisuperspace model, 
at least as a starting exercise towards tackling more involved and general
scenarios. 

We conclude with the modest statement that the analysis in theory space,
so far, has not yielded any startling new results. The geometries obtained
in various contexts as well as the result about the nature of RG flows
discussed in the previous section necessarily restate known results in
a different language. Morever, the construction of the line element
is certainly dependent on the integrability or renormalisability of the
field theory. A better understanding of the merits/demerits of this
approach would require more worked-out examples. Perhaps the central issue
to address would be `What are the distinct features which can be obtained  
{\em exclusively} from a theory space analysis ?'. A clear answer to this
query will demonstrate the power, if any, of this approach, in future.

\noindent

\nonumsection{Acknowledgments}
\noindent
The author thanks D.V. Ahluwalia and N. Dadhich for inviting him to
speak in the Workshop on Interface of Gravitational and Quantum 
Realms (IGQR-I) and also for the warm hospitality extended by
IUCAA, Pune during the workshop.

\nonumsection{References}

\end{document}